\documentclass[reprint,aps,prd,superscriptaddress,showpacs,preprintnumbers,nofootinbib]{revtex4-2}
\usepackage{amssymb, graphics, setspace}
\usepackage{pgfplots}
\usepackage{textcomp}
\usepackage[caption=false]{subfig}
\usepackage{amsmath}
\usepackage{commath}

\usepackage{graphicx,bm}
\usepackage{url}
\usepackage{float}
\usepackage{textcomp}
\usepackage{verbatim}
\begin{document}
	
	\newcommand{\dlt}{\bigtriangleup}
	\newcommand{\beq}{\begin{equation}}
		\newcommand{\eeq}[1]{\label{#1} \end{equation}}
	\newcommand{\insertplot}[1]{\centerline{\psfig{figure={#1},width=14.5cm}}}
	
	\parskip=0.3cm
	
	
	\title{Nonlinear Regge trajectories and saturation of the Hagedorn
spectrum}
	
	
	\author{Istv\'an Szanyi}
	\email{szanyi.istvan@wigner.hu}
	\affiliation{Wigner Research Centre for Physics, H-1525 Budapest 114, POBox 49, Hungary}
	\affiliation{E\"otv\"os University, H-1117 Budapest, P\'azm\'any P. s. 1/A, Hungary}
	\affiliation{MATE Institute of Technology,  K\'aroly R\'obert Campus, H-3200 Gy\"ongy\"os, M\'atrai \'ut 36, Hungary}
	
	\author{Tam\'as Bir\'o}
	\email{biro.tamas@wigner.hu}
	\affiliation{Wigner Research Centre for Physics, H-1525 Budapest 114, POBox 49, Hungary}

	\author{L\'aszl\'o Jenkovszky}
	\email{jenk@bitp.kiev.ua}
	\affiliation{Bogolyubov Institute for Theoretical Physics (BITP),
		Ukrainian National Academy of Sciences \\14-b, Metrologicheskaya str.,
		Kiev, 03680, Ukraine}

	\author{Vladyslav Libov}
	\email{vladyslav.libov@gmail.com}
	\affiliation{Bogolyubov Institute for Theoretical Physics (BITP),
		Ukrainian National Academy of Sciences \\14-b, Metrologicheskaya str.,
		Kiev, 03680, Ukraine}

	\begin{abstract}
		We argue that two seemingly different phenomena, namely the well-known saturation of the Hagedorn exponential distribution and the less familiar saturation of Regge trajectories at resonance masses $m\approx$ 2--2.5 GeV are related and have the same origin: quark deconfinement. We show that the slope of the real part of non-linear 
		Regge trajectories determines the prefactor $f(m)$ in Hagedorn's 
		resonance mass density distribution $\rho(m)$. While the Hagedorn distribution comes from statistics, Regge trajectories contain dynamics. 
		
	\end{abstract}
	
	
	\maketitle
	
	\section{Introduction}
	
	\noindent The spectrum of hadron resonances is among the central problems of high-energy physics. The properties of highly excited resonances are intimately connected with the problem of confinement. We revise the spectrum of hadronic resonances by relating seemingly two different viewpoints: statistical (Hagedorn) and dynamical (Regge).
	
	The density of hadron states follows the law \cite{H}
	\begin{equation}\label{Eq:H}
		\rho(m)=f(m)\exp(m/T_H),
	\end{equation}    
	where $f(m)$ is a slowly varying function of mass and $T_H$ is the Hagedorn temperature, 
	originally considered as the limiting temperature but later interpreted as the temperature 
	of the color deconfinement phase transition where hadrons "boil" transforming 
	the matter into a boiling quark-gluon soup.
	
	Over 50 years after the publication of R. Hagedorn's paper \cite{H} on the spectrum of resonances many important details still remain open. In spite of many efforts by different groups of authors~\cite{Bron_00,Bron_F,Cohen, Cleymans} involving more new data on resonances, the calculated value of the Hagedorn temperature shows a surprisingly widespread from $T_H=141$ MeV to $T_H=340$ MeV, depending on the parametrization and the set of data (baryons, mesons) used. The discrepancies may have different origin, in particular: a) the large uncertainties in the specification and identification of heavy resonances, b) the analytical form of the Hagedorn spectrum, in particular, the form of the function $f(m)$ multiplying Hagedorn's exponential. In the present paper, we address both issues.  
	
	In our analyses, we rely on the latest results of the Particle Data Group \cite{Workman:2022ynf}. At the same time, we are aware of recent important developments and progress in theoretical studies of the Hagedorn spectrum and of the ultimate temperature based on gauge theories~\cite{Bigazzi2022},  lattice QCD (by which one distinguishes between two different types of resonances: isolated ones in vacuum, and resonance at a certain temperature), strings, supergravity~\cite{Bringoltz2006,Caselle2015,Kutasov1991}. Its possible manifestation in the early universe was discussed in Refs.~\cite{Huang1970,Tye1985,Atick1988}.
	
	Our approach, however, is limited to the world of observed
	resonances, summarized by the Particle Data Group and theoretical methods based on analyticity, unitarity, and duality.
	
	Crucial and original in our paper is the identification of the function $f(m)$ with the derivative of the relevant Regge trajectory. In the spirit of the analytic $S$-matrix approach, Regge trajectories encode an essential part of the strong interaction dynamics, they are building blocks of the theory. There were many attempts to find analytic forms of the non-linear complex Regge trajectories, based on mechanical analogs (string), quantum chromodynamics etc., suggesting different and approximate solutions applicable in a limited range. In our approach, we rely on duality and constraints based on analyticity and unitarity, constraining the threshold and asymptotic behavior of the trajectories. An important constraint, affecting the spectrum near its critical point is an upper bound on the real part of Regge trajectories', coming from dual models with Mandelstam analyticity \cite{Bugrij1973}. Construction of explicit models of the trajectories satisfying the above constraints is a non-trivial problem. In the present paper, we propose explicit models of such trajectories allowing explicit calculations and compatible with the data on resonances.

	Below we argue that while Hagedorn's exponential rise comes mainly from the proliferation of spin and isospin degeneracy of states with increasing mass, that can be counted directly, the prefactor $f(m)$ reflects dynamics, encoded in Regge trajectories given that $f(m)=\alpha'(m)$, where $\alpha'(m)$ is the slope of the trajectory.

	The low-mass, $m<1.8$ GeV spectra do not exhibit any surprise by following Hagedorn's exponential. The  only open questions are the value of the Hagedorn temperature $T_H$ and possible differences between the spectra for various particles. The spectra beyond $m=1.8$ GeV are different: on the experimental side, the high-mass resonances tend to gradually disappear, their status becoming uncertain. An update of the Hagedorn mass spectrum can be found in \cite{Bron_00,Bron_F, Cohen, Cleymans}. Different Hagedorn temperatures for mesons and baryons from experimental mass spectra, compound hadrons, and combinatorial saturation were studied in \cite{Bron_00}. 
	
	In any case, the most important issue is the existence of a ``melting point", where the resonances are transformed to a continuum of a boiling soup of quarks and gluons. This critical region/point is studied by various methods: statistics and thermodynamics, quantum chromodynamics, lattice quantum chromodynamics (QCD) calculations, the MIT bag models, Regge poles and relations derived within analytic $S$ matrix theory. The research objectives are the order of the assumed phase transition or crossover transition,  fluctuations and correlations of conserved charges and the Fourier coefficients of net-baryon density. For a recent treatment of these issues see Ref. \cite{Goren}.
	
	In the present paper, we extend the existing panorama by appending the dynamics arising from the behavior of Regge trajectories. Regge trajectories are rich objects containing information on the spectrum of resonances. Although they are usually approximated by infinitely rising linear functions, predicting an infinite number of resonances, in fact, the analytic theory confines the rise of their real part limiting the  number of resonances in Nature.  Below we show how the nonlinear complex trajectories affect the Hagedorn spectrum.          
	
   More details on the Hagedorn spectrum, Hagedorn temperature, and hot phases of hadronic matter can be found in the writings of Johann Rafelski \cite{R1, Rafelski, R2}.

	The paper is organized as follows. In  Subsec. \ref{Hagedorn} we discuss various approaches to the density of hadron states, including the Hagedorn model. In Subsec. \ref{Regge} we discuss complex, non-linear Regge trajectories and their relation to Hagedorn's density of states. Both, the density of states and the hadron spectrum are finite and interrelated, as shown in Subsec. \ref{rho} by a relevant explicit example. In Sec. \ref{Boiling} we discuss the relation between the spectra and the statistical properties of the nuclear matter with possible equations of state of hadronic matter.

	\section{Melting hadrons}
	
	The spectrum of resonances and their statistical properties are interrelated. In this section we study the relation between the mass density of hadronic states given by the Hagedorn spectrum and the dynamics emerging from Regge pole models, inspecting  non-linear Regge trajectories. 
	
	In spite of the huge number of papers, the  subject remains a topical problem of hadron dynamics with numerous open questions. In this paper we address the following issues: 
	\begin{itemize} 
		\item the behavior of the  meson mass spectrum in the high-mass region;
		\item the role of the critical temperature and the prefactor in $\rho(m)$ in the Hagedorn model of hadronic spectra;
		\item the finiteness of the Hagedorn spectrum and its consequences.
	\end{itemize}

	\subsection{Density of states (Hagedorn distribution) and resonance spectra (Regge trajectories)} \label{Hagedorn}
	
	The idea of the spectral description of a strongly interacting gas was suggested  by 
	Belenky and Landau \cite{Landau}. This approach treats  resonances on equal footing 
	with stable hadrons. The expression for pressure in this thermodynamic approach 
	in the Boltzmann approximation is given by: 

	\begin{equation}
		p=\sum_ig_ip(m_i)=\int_{M_1}^{M_2}\limits\! dm\, \rho(m) \, p(m),
	\end{equation}
	with
	\begin{equation}\nonumber
		p(m)=\frac{T^2m^2}{2\pi^2}K_2\bigl(\frac{m}{T}\bigr),
	\end{equation}
	where $M_1$ and $M_2$ are the masses of the lightest and heaviest hadrons, 
	respectively, and $g_i$-s are particle degenerations. 
	
	It was suggested in 
	Refs. \cite{Shuryak1, Shuryak, Burak2} that for fixed isospin and hypercharge a cubic density 
	of states, $\rho(m)\sim m^3,$ fits the data. Moreover, as argued in Ref. \cite{Shuryak1}, 
	the cubic spectrum can be related to collinear Regge trajectories. 
	Indeed, following the arguments of Burakovsky \cite{Burak}, on a linear trajectory 
	with negative intercept, $\alpha(t)=\alpha' t-1$, some integer values of $\alpha(t)=J$ 
	correspond to states with negative spin, $J=\alpha(t_J),$ 
	with squared masses $m^2(J)=t_J.$  
	Since a spin-$J$ state has multiplicity $2J+1$, 
	the total number of states with spin $0\leq J\leq j$ at $t=m(j)^2$ is given by
	\begin{equation}
		N(j)=\sum_{J=0}^j(2J+1)=(j+1)^2=\alpha'^2m^4(j).
	\end{equation}
	Hence the density of states per unit mass interval is obtained as the derivative
	of this cumulative quantity,
	\begin{equation}
		\rho(m)=\frac{dN(m)}{dm}=4\alpha'^2m^3,
	\end{equation}
	and it grows as the cubic power of the mass.
	Consequently for a finite number of collinear trajectories, $N$, 
	the corresponding mass spectrum is given as
	\begin{equation}
		\rho(m)=4N\alpha'^2m^3.
	\end{equation}
	
	A different view on the spectra was advocated by Shuryak \cite{Shuryak}, 
	who suggested to use a quadratic parametrization, 
	completely different from the conventional form:
	$$\rho(m)\sim m^2.$$
	
	

	In both the statistical bootstrap model \cite{H} and in the dual resonance model \cite{dual, Statboot} the resonance spectrum takes the form of Eq.~(1). In the dual resonance model $f(m)\sim \frac{d}{dm}\mathfrak{Re}\alpha(m^2).$ In this work  we use non-linear complex Regge trajectories to determine this prefactor as discussed in the next subsections.
	
	The meson and baryon spectra differ, in particular by their slopes, 
	as shown e.g. in Refs.~\cite{Bron_00, Bron_F}.  
	More important is the question of the asymptotic behavior of $\rho(m)$
	for large masses. 
	In theory, Hagedorn's exponential may rise indefinitely, however, starting from $m\approx 2.5~\textrm{GeV}$ resonances are not observed.
	The question arises whether it is a "technical" issue 
	(the resonances gradually fade becoming too wide  to be detected) or there is a critical point 
	where they melt to a continuum transforming the hadron matter to a "boiling soup". 
	Remarkably this point can be illuminated by means of Regge trajectories, 
	as we shall demonstrate it in the following. 
	
	In the present paper, we concentrate on the meson spectrum, more specifically that of $\rho$ and its excitations. This familiar trajectory is chosen just as a representative example. Other trajectories, including baryonic ones as well as those with heavy ($c$ and $b$) flavors will be studied later. We are interested in the high-mass behavior, starting from $m\approx1.8$ GeV. Beyond this value the exponential behavior of the Hagedorn spectrum is expected to change drastically. Referring to perfect low- and intermediate mass fits of  \cite{Bron_F}, we concentrate now on its behavior above $1.8$ GeV. 
	
	
	Note that instead of comparing the density of states $\rho(m)$ to the data it is customary to accumulate states of masses lower than $m$,
	\begin{equation} \label{Eq:density} N_{exp}=\sum_ig_i\Theta(m-m_i),
	\end{equation}
	where $g_i$ is the degeneracy of the i-th state with mass $m_i$ in spin $J$ and isotopical spin $I$, i.e., 
	\[
	g_i=
	\Bigg\{\begin{array}{ll}
		(2J_i+1)(2I_i+1), &  \textrm{for non-strange mesons} \\
		4(2J_i+1), &  \textrm{for strange mesons} \\
		2(2J_i+1)(2I_i+1), &  \textrm{for baryons} 
	\end{array}
	\]
	The theoretical equivalent of Eq. (\ref{Eq:density}) is \begin{equation}\label{Eq:density1} N_{theor}(m)=\int_{m_\pi}^m\rho(m')dm', \end{equation}
	where the lower integration limit is given by the mass of the pion.
	We  identify $f(m)$ in $\rho(m)$ with the slope of the relevant non-linear complex Regge trajectory $\alpha'(m)$. In the next subsection we discuss the properties of these trajectories 
	following from the analytic $S-$matrix theory and duality, and present an explicit example of such a trajectory.    
	
	\subsection{Regge trajectories}\label{Regge}
	
	At low and intermediate masses, light hadrons fit linear Regge trajectories with 
	a universal slope, $\alpha'\approx 0.85$ GeV$^2$. As masses increase, the spectrum 
	changes: resonances tend to disappear. The origin and details of this change are disputable.
	
	Termination of resonances, associated with a "ionization point" was also studied  
	in a different class of dual models, based on logarithmic trajectories \cite{Coon}. 
	
	Possible links between the Hagedorn spectra and Regge trajectories appear in the statistical bootstrap and dual models \cite{Statboot}, according to which the prefactor $f(m)$ in Eq. (\ref{Eq:H}) depends on the slope of the relevant Regge trajectory, $\alpha'(m^2)$, which is a constant for linear Regge trajectories.
	
\begin{figure}[ht] 
		\center{\includegraphics[width=.46\textwidth]
			{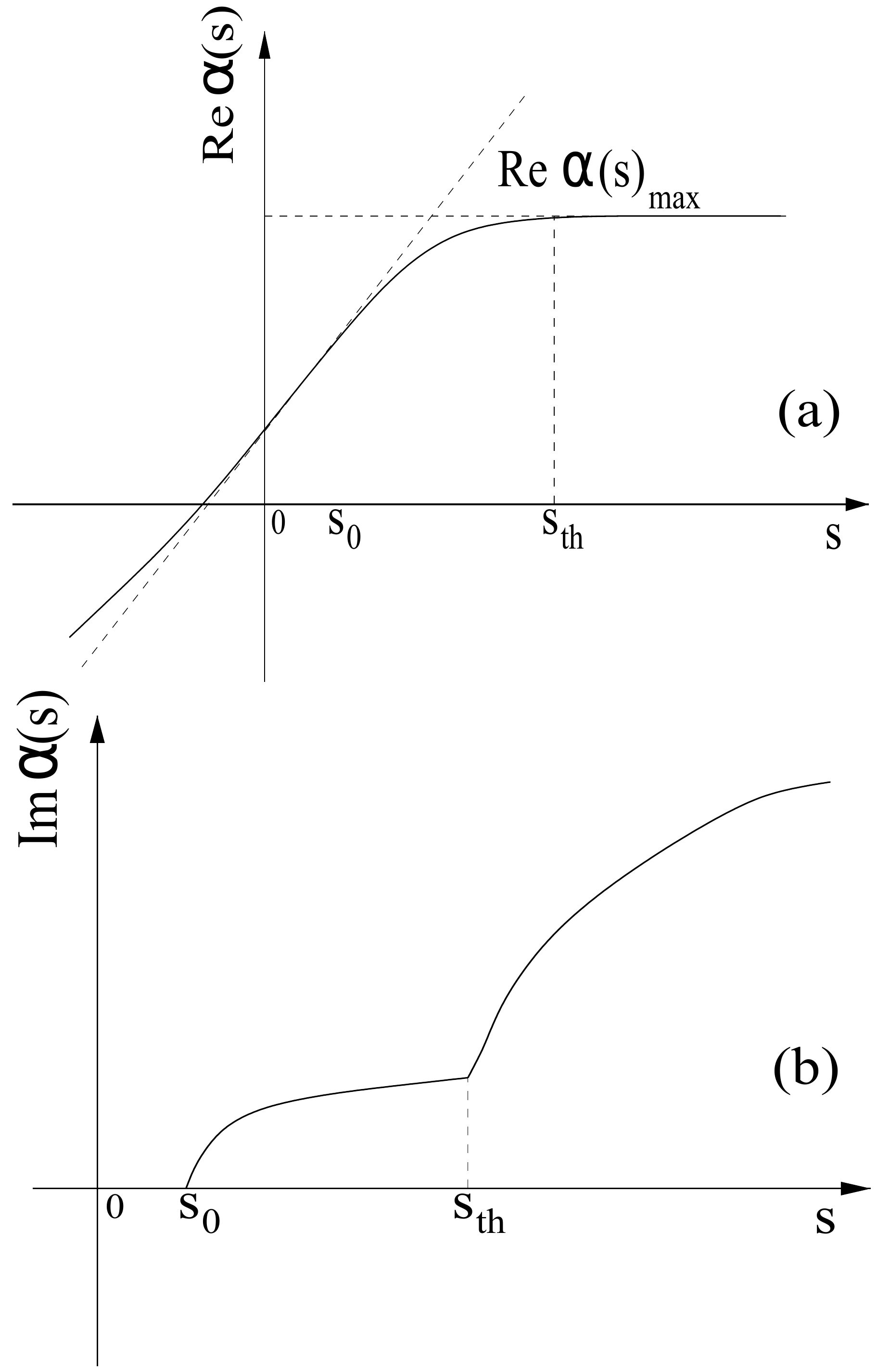} }
		\caption{Typical behavior of the real (a) and imaginary (b) parts of Regge trajectories in dual models with Mandelstam analyticity \cite{Bugrij1973}.}
		\label{Fig:alpha}
	\end{figure}

	We extend the Hagedorn model
	by introducing the slope of 
	relevant non-linear Regge trajectories. Anticipating a detailed quantitative analysis, 
	one may observe immediately that a flattening of 
	$\mathfrak{Re}\alpha(s=m^2)$ \footnote{We use the (here positive) variables $s$ or $t$ interchangeably 
		with crossing-symmetry in mind.}, shown in Fig.~\ref{Fig:alpha} , results in a 
	decrease of the relevant slope $\alpha'(m)$ and a corresponding change 
	in the Hagedorn spectrum. Following Eq. (\ref{Eq:H}) we parametrize 
	\begin{equation}\label{Eq:H1} \rho(m)=\left(\frac{d}{dm}\mathfrak{Re}\alpha(m^2)\right)\times \exp(m/T_H). 
	\end{equation} 
	Based on the decreasing factor $\mathfrak{Re} \alpha^{\prime}$ in Eq.~(\ref{Eq:H1}) the exponential rise of the density of states slows down near to the melting point around $ m \approx 2 -2.5$ GeV. The cumulative spectrum Eqs. (\ref{Eq:density}) and (\ref{Eq:density1}), accordingly  tends to a constant value.  
	
	Any Regge trajectory should satisfy the followings (for a comprehensive review see~\cite{Trush}):
	\begin{itemize}
		\item threshold behavior imposed by unitarity;
		\item asymptotic  constraints: the rise of real part of Regge trajectories is limited, $\Re\alpha(s)\leq\gamma\sqrt{t}\ln t,\ \  s\to\infty$; 
		\item compatibility with the nearly linear behavior in the resonance 
		region (Chew-Frautschi plot).
	\end{itemize}
	
	The threshold 
	behavior of Regge trajectories is constrained by  unitarity. As shown by Barut and Zwanziger 
	\cite{Barut}, $t$-channel unitarity constrains the Regge trajectories near 
	the threshold, $t\rightarrow t_0$  to the form
	\begin{equation} \label{Eq:Barut}
		\mathfrak{Im} \alpha(t)\sim (t-t_0)^{\mathfrak{Re}\alpha(t_0)+1/2}.
	\end{equation} 
	Here $t_0$ is the lightest threshold, {\it e.g.} $4m_{\pi}^2$ 
	for the meson trajectories. Since $\mathfrak{Re}\alpha(4m_{\pi}^2)$ is small, 
	a square-root threshold is a reasonable approximation to the above constraint. 
	
	In the resonance region below flattening near
	$m=\sqrt{s}\lesssim 2.5$ MeV the meson and baryon 
	trajectories are nearly linear (Chew-Frautschi plot). 
	Fixed-angle scaling behavior of the amplitude constrains the trajectories 
	even more, down to a logarithmic behavior \cite{BCJ}.

	The combination of the threshold behavior Eq. (\ref{Eq:Barut}) and the above asymptotic behavior suggested the explicit model \cite{BK} of trajectories as a sum of square-root thresholds (see the details in Subsec.~\ref{rho}).
	
	There are several reasons why the non-linear and complex nature of the Regge 
	trajectories is often ignored, namely:
	1) the observed spectrum of meson and baryon resonances (Chew-Frautschi plot) seem to 
	confirm their linearity; 2) in the scattering region, $t<0,$ the differential 
	cross-section, $d\sigma/dt\sim\exp((2\alpha(t)-2) \ln s)$ is nearly exponential in $t$;
	3) Dual models, {\it e.g.} the Veneziano amplitude are valid only in the 
	narrow-width approximation, corresponding to linear Regge trajectories (hadronic strings). Deviation from linearity is unavoidable, but its practical realization is not easy. Attempts are known in the literature, see e.g. \cite{ Paccanoni_M, Paccanoni_B, Burak} and references therein. 
	
	As a final remark, we comment on
	a typical feature of dual analytic models, namely the extremely broad resonance approximation, suggested in~\cite{Bugrij1973} by which the unitarization procedure, contrary to the narrow resonance (Veneziano-like) models Fig. 25 of Ref.~\cite{Bugrij1973}, move the resonance pole from the real (negative) axis to the physical region of broad resonances, harmless to its basic properties, e.g. polynomial boundedness (for more details see~\cite{Bugrij1973}). This property is essential for the parametrization of our Regge trajectory $\alpha(t)$. Due to our factor $f(m)=(Re\alpha(m))’$, the spectrum is a “mixture” of narrow and wide resonances, therefore not purely “Hagedornian” anymore.  
	
	\subsection{The $\rho$-trajectory and the Hagedorn spectrum} \label{rho}
	\noindent
	Trajectories satisfying
	the above requirements have been studied extensively in the past. 
	A particularly simple and transparent non-linear trajectory was suggested in Refs. \cite{BK, BK1} and is defined as a sum of square-root thresholds formed by stable particles, allowed by quantum numbers. While the imaginary part of such a trajectory starts to be nonzero by exceeding the lightest threshold and rises indefinitely, its real part terminates at the heaviest threshold (see Fig.~\ref{Fig:alpha}). 
	
	Specifically, Ref.~\cite{BK1} defines trajectories as
	\begin{equation}\label{eq:bk_rho}
		\alpha(s)=\lambda-\sum_i \gamma_i \sqrt{s_i-s},
	\end{equation}
	where all two-particle stable thresholds are included in the sum.
	For the $\rho$-meson trajectory two meson-meson and four baryon-antibaryon channels 
	are taken into account, as listed in Table~\ref{tab:rho_channels} along with corresponding 
	thresholds $s_i$ and weights $\gamma_i$.
	Note, that for simplicity the same weight is used for all baryon-antibaryon channels.
	\begin{table}[h]
		\centering
		\begin{tabular}{c|c|c}
			Channel & $s_i$ & $\gamma_i$\\
			\hline
			$\pi\pi$ & 0.078 & 0.127\\
			$K\overline{K}$ & 0.976& 0.093   \\
			$N\overline{N}$ &  3.52 & 0.761\\
			$\Lambda\overline{\Sigma}$ &  5.31 & 0.761 \\
			$\Sigma\overline{\Sigma}$ &  5.66 & 0.761\\
			$\Xi\overline{\Xi}$ &  6.98 & 0.761\\
			
		\end{tabular}
		\caption{Parametrisation of the $\rho$-trajectory~\cite{BK1}.}
		\label{tab:rho_channels}
	\end{table}
	
\begin{figure}[hbt]
		\center{\includegraphics[width=.52\textwidth]{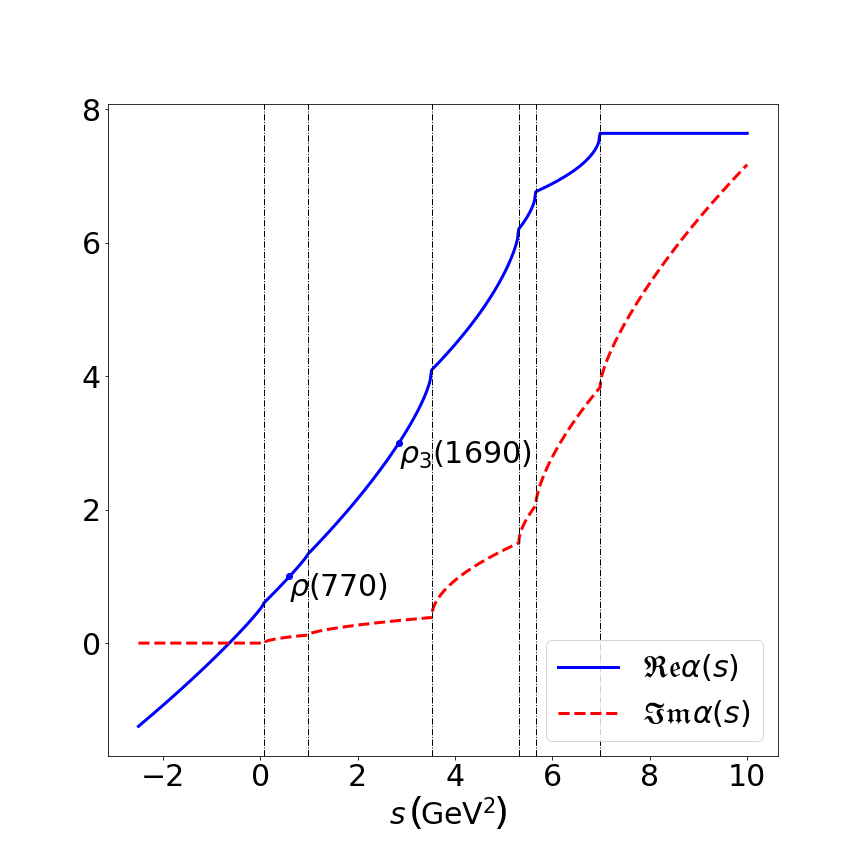}}
		\caption{Real and imaginary parts of the $\rho$-trajectory.}
		\label{fig:rho_alpha}
	\end{figure}

	The real and imaginary parts of the trajectory described by Eq.~(\ref{eq:bk_rho}) with parameters from Table~\ref{tab:rho_channels}
	are shown in Figure~\ref{fig:rho_alpha}. The vertical lines indicate thresholds $s_i$.
	As expected, above the highest threshold ($s_6=6.98~\textrm{GeV}^2$) the real part reduces to a
	constant, $\mathfrak{Re}\alpha=\lambda$.
	
	\begin{figure}[ht]
		\center{\includegraphics[width=.52\textwidth]{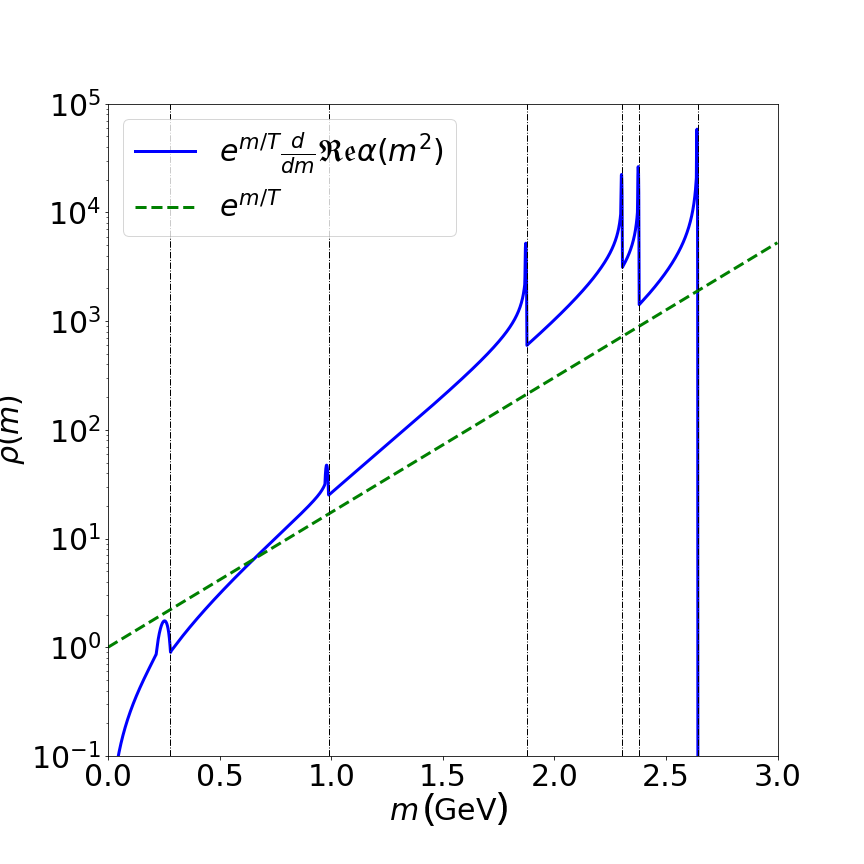}}
		\caption{Mass density $\rho(m)$ calculated by using the derivative of the smoothed $\rho$-meson trajectory as a prefactor, Eq.~(\ref{Eq:H1_rho}).
			The exponential density without any prefactor is also shown (dashed line), using the same temperature and normalization. Above the highest threshold the derivative, and hence the density vanishes (not seen due to the logarithmic scale).}
		\label{fig:density}
	\end{figure}
	
	Putting $s=m^2$ and differentiating the real part of the trajectory with respect to $m$ we obtain
	\begin{equation}\label{eq:d_dm_re}
		\frac{d}{dm}\mathfrak{Re}\alpha(m^2)=\sum_{i} \frac{\gamma_im}{\sqrt{s_i-m^2}},
	\end{equation}
	where for a given $m$ the sum includes only terms satisfying $m^2<s_i$.
	Substitution into (\ref{Eq:H1}) yields
	\begin{align}\label{Eq:H1_rho}
		\rho(m) =
		\frac{d}{dm} \mathfrak{Re} \alpha(m^2) \exp(m/T) \\ \nonumber =
		\sum_{i:\,s_i>m^2} \frac{\gamma_i m}{\sqrt{s_i-m^2}} \exp(m/T).
	\end{align}
	
	As in Eq.~(\ref{eq:d_dm_re}), the sum includes for a given $m$ only terms with $s_i>m^2$;
	the derivative and thus the mass density vanish above the highest threshold.
	At the thresholds $s_i$, the derivative and hence the density diverge. However, the relevant integral, Eq.~(\ref{Eq:density1}) stays finite. For simplicity, we smooth the trajectory $\mathfrak{Re}\,\alpha(s)$ in the vicinity of the thresholds using splines.
	With this procedure the derivative becomes finite and continuous, and can be integrated numerically.
	An example of the resulting mass density $\rho(m)$ is shown in Figure~\ref{fig:density}.
	
\begin{figure}[ht]
		\center{\includegraphics[width=.52\textwidth]{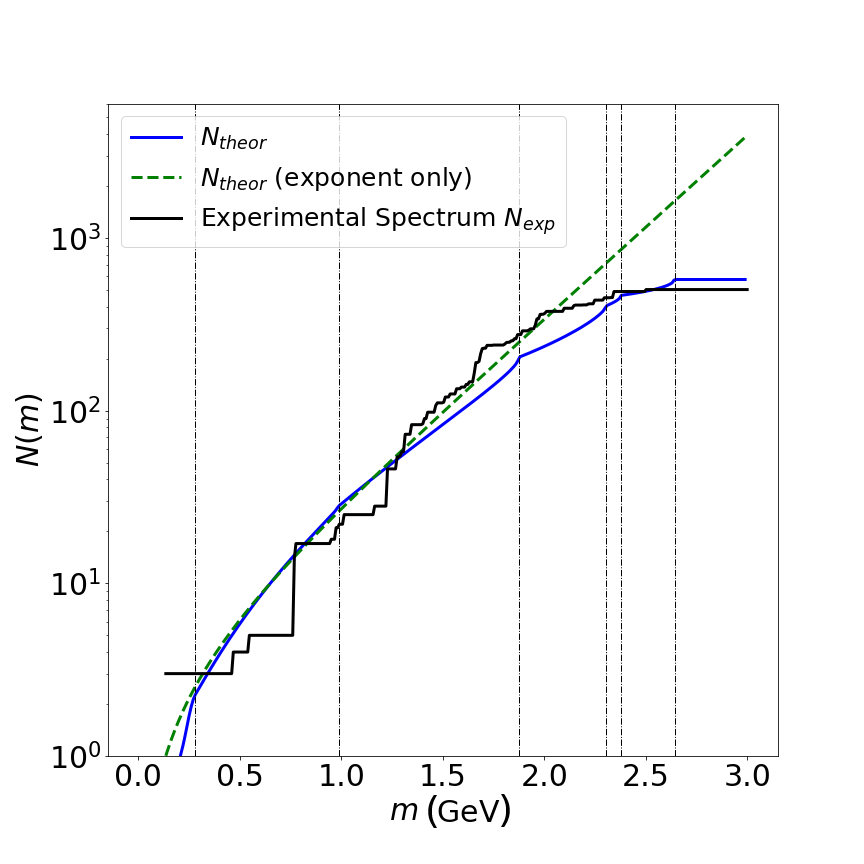}}
		\caption{Hadron mass spectrum $N_{theor}$ from Eq.~(\ref{Eq:density1}) using the mass density of Eq.~(\ref{Eq:H1_rho}) (blue line) compared to data (black histogram). Additionally, $N_{theor}$ obtained from a simple exponential mass density is also shown (green dashed line).} 
		\label{fig:mass_spectrum}
	\end{figure}

	The mass density $\rho(m)$ can now be integrated using Eq.~(\ref{Eq:density1}) in order to obtain the theoretical prediction for the mass spectrum, $N_{theor}(m)$. 
	Figure~\ref{fig:mass_spectrum} shows the result, together with the experimental cumulative spectrum $N_{exp}(m)$ using the states listed by the Particle Data Group~\cite{Workman:2022ynf}. 
	
	As expected, $N_{theor}(m)$ flattens above the highest threshold (corresponding to $m\sim 2.65~\textrm{GeV}$), since the density $\rho(m)$ defined with Eq.~(\ref{Eq:H1_rho}) vanishes in that region. This is consistent with the experiment, since no unflavored mesons have been observed above $m\sim 2.5~\textrm{GeV}$~\cite{Workman:2022ynf}.
	On the other hand, the integrated spectrum using a simple form $\rho=e^{m/T}$ is rising indefinitely, thus failing to describe flattening of the data at high masses.
	
	The temperature and normalization used in Fig.~\ref{fig:mass_spectrum} were determined from a least-square fit to data. The fit was performed separately for two models (a simple exponent and an exponent with a prefactor). The optimal temperature for the pure exponent is $T=0.41~\textrm{GeV}$, slightly larger but generally consistent with the previous studies. Inclusion of the prefactor steepens the curve (see e.g. Fig.~\ref{fig:density}), thus a larger temperature ($T=1.45~\textrm{GeV}$) is required to fit the data.

	Resonances may melt for two reasons: 
	1) for square-root $\alpha(s)$ trajectories (\ref{eq:bk_rho}), 
	whose real part terminates at some large mass and consequently the derivative vanishes, 
	or 2) for densities of resonances modelled by power-like functions 
	rather than exponentials.
	
	The above is only a representative example intended to show the interrelation between the Hagedorn spectrum and the one based on non-linear complex Regge trajectories (Chew-Frautchi plot). The present results can be extended in several directions:
	\begin{itemize}
		\item by including other trajectories - mesonic and/or baryonic;
		\item by including heavy flavors such as $b$ quarks,     
		both in the Hagedorn and Regge spectra.
		\item working with alternative  parametrizations of non-linear, complex Regge trajectories. 
	\end{itemize}

	\section{Boiling quarks and gluons} \label{Boiling}
	In the previous section, we inspected the spectrum of resonances by combining two different approaches - statistical (Hagedorn) and dynamical (Regge). We have focused on the region of heaviest resonances, the region where hadrons may melt transforming in a boiling "soup" of quarks and gluons. Melting may happen in different ways, characterized by the details for a phase transition of colorless hadronic states into a quark-gluon soup, whatever it be \cite{BJSch}. In terms of hadron strings this process corresponds to breakdown (fragmentation) of a string. Lacking any theory of confinement providing a quantitative description of interacting string, we will not pursue this model. Instead we use thermodynamics adequate in this situation.  To complement the previous section, we present  our arguments below related to the possible change of phase from a different, thermodynamic perspective.

	The Hagedorn exponential spectrum of resonances Eq. (\ref{Eq:H}) results in 
	a singularity in the thermodynamic functions at critical temperature $T=T_c$  
	and in an infinite number of effective degrees of freedom in the hadronic phase. 
	Also, as shown in Ref. \cite{Fowler}, a Hagedorn-like mass spectrum is incompatible 
	with the existence of the quark-gluon phase. To form a quark phase from the hadronic phase, 
	the hadron spectrum cannot grow more quickly than a power. This is possible \cite{Shuryak} in 
	case of a simple power parametrization $\rho\sim m^k,$ compatible with $k\approx 3,$ 
	for the observable mass spectrum in the interval 0.2--1.5 GeV. 
	Assuming ideal contributions to thermodynamical quantities we hence take 
	energy density in the form
	\begin{equation} \label{ENDENS5}
		\epsilon \: = \: \int_0^{\infty}\limits \rho(m) T^4 \sigma(m/T) dm
		\: = \: \lambda_k T^{k+5},
	\end{equation}
	and obtain the corresponding pressure and sound velocity square as follows\footnote{Note that the definition of entropy density $s$, energy density $\epsilon$ and velocity of sound $c_s$ in case of $\mu=0$: $$s(T)=p'(T),\ \epsilon(T)=Ts-p,\ \ c_s^2=\frac{dp}{d\epsilon}=\frac{p'}{Tp''}=\frac{s}{Ts'}.$$}:
	\begin{equation} \label{PRESSURE}
		p=\frac{\lambda_k}{k+4}T^{k+5},\ \ c_s^2=1/(k+4).
	\end{equation}
	
	The above empirical power-like behavior has also theoretical background. 
	In Ref. \cite{Jenkovszky} an asymptotic, $T\gg m$, EOS was derived by using 
	the $S$-matrix formulation of statistical mechanics \cite{DMB}. 
	It was shown that the existence of the forward cone in hadronic interactions 
	with non-decreasing total cross sections, $i.e$, pomeron dominance, confirmed by numerous experiments 
	at high energies, results in an asymptotic, $T\gg m$ EOS $p(T)\sim T^6$ where 
	$m$ is a characteristic hadron (e.g. pion) mass. The inclusion of non-asymptotic (secondary) Regge terms produces a minimum in the $p(T)$ dependence at negative pressure, with far-reaching observable consequences.  
	
	Such an asymptotic formula for the pressure, different from the "standard" $p(T)\sim T^4$, 
	was derived, on different grounds also in Ref. \cite{Shuryak}.

	The unorthodox $p\sim T^6$ asymptotic behavior is orthogonal to the "canonical" (perturbative QCD) form $\sim T^4$. Still, it cannot be rejected e.g. when assuming a screening of the action of large-distance van der Waals forces at high temperatures and densities. 
	
	In Ref. \cite{Bugrij} the asymptotic form $\sim T^6$ was extended to lower 
	temperatures by adding non-asymptotic Regge-pole exchanges. The resulting EOS is 
	\begin{equation}
		p(T)=aT^4-bT^5+cT^6,
	\end{equation}
	where $a,\ b,\ c$ are parameters connected with Regge-pole fits to high-energy 
	hadron scattering. The remarkable property of this EOS, apart from the non-standard 
	asymptotic behavior, $\sim T^6,$ is the appearance of the non-asymptotic 
	term $T^5$ with negative sign, creating a local minimum with negative pressure. 
	This metastable state with negative pressure was shown \cite{inflate} to produce 
	inflation of the universe. 
	
	In the next part, we discuss the phase transition of colorless hadronic states into quark-gluon soup in the framework of modified quark bag models. 
	
\begin{figure*}[ht]
		\centering
		\includegraphics[width=.8\textwidth]{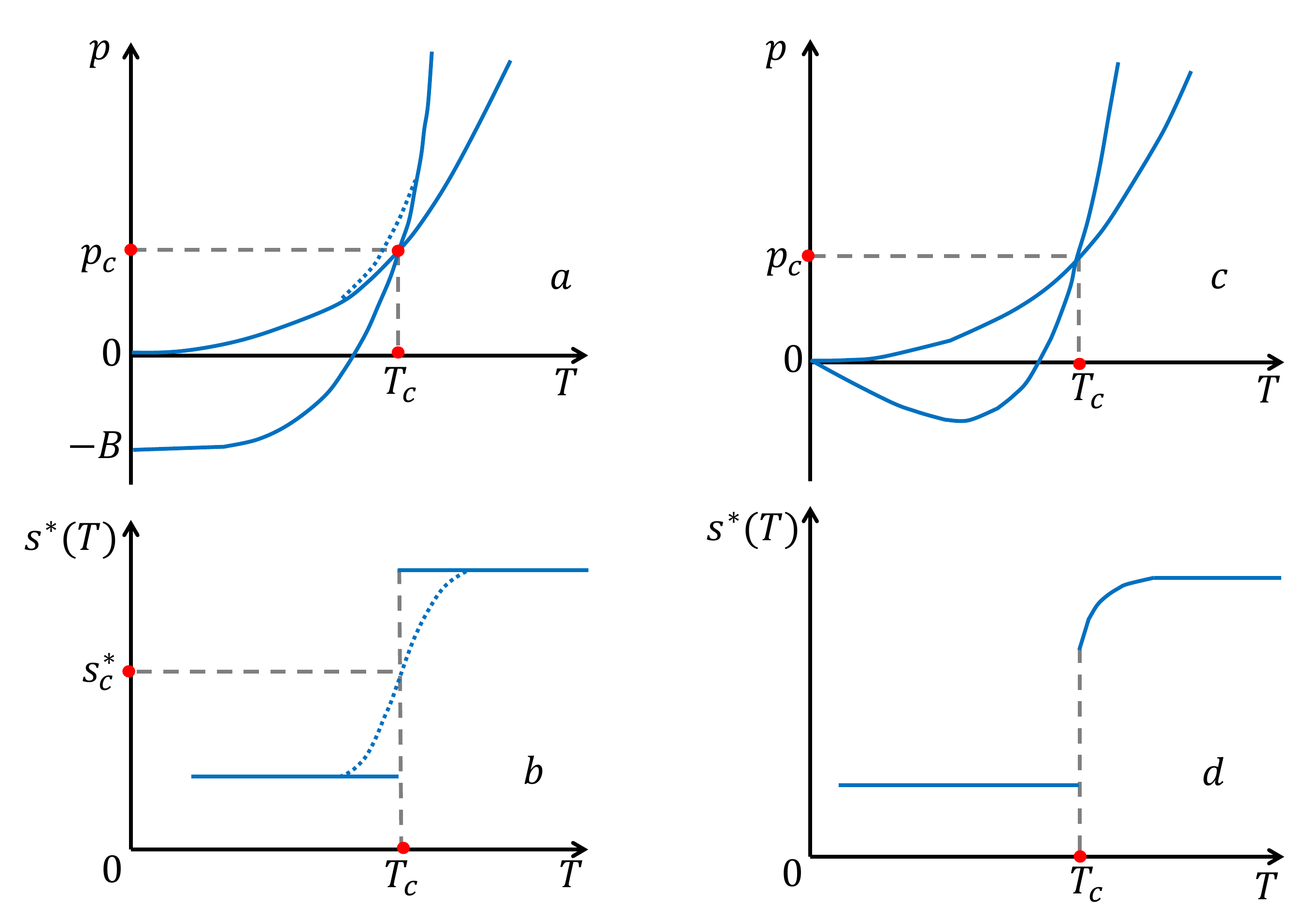}
		\caption{Icons $a$ and $b$ show the standard bag EoS, Eq. (\ref{Eq:Bag}), with constant $B$; here $s^*(T)=s(T)/T^3.$ The dotted line corresponds to the modified bag EoS (\ref{Eq:EOS1}).
			Icons $c$ and $d$ feature the K\"allmann EoS \cite{Kallmann}, where $B=AT$.\label{Fig:6}}
	\end{figure*}

	The standard bag equation of state assuming, for simplicity, vanishing chemical potential, $\mu=0$:
	\begin{equation}\label{Eq:q}
		p_q(T)=\frac{\pi^2}{90}\nu_qT^4-B,
	\end{equation}
	\begin{equation}\label{Eq:Bag}
		p_h(T)=\frac{\pi^2}{90}\nu_hT^4,
	\end{equation}
	where $p_q(T)$ and $p_h(T)$ are pressure in the quark-gluon plasma (QGP) and in the hadronic gas phase, respectively, $B$ is the bag constant, and $\nu_{q(h)}$ is the number of degrees of freedom in the QGP (hadronic gas). 
	From Eqs. (\ref{Eq:q}) and (\ref{Eq:Bag}) one finds the characteristic parameters of the phase transition (see Fig.~\ref{Fig:6}a):
	\begin{equation}
		p_c=B\nu_h/(\nu_q-\nu_h),\ \ T_c=[90B/\pi^2(\nu_q-\nu_h)]^{1/4}.
	\end{equation}
	Since $s(T)=dp(T)/dT,$ the relevant formula for 
	the entropy density can be rewritten as (see Fig.~\ref{Fig:6}b) 
	\begin{equation}\label{eq:EOS_ent1}
		s(T)=(2\pi^2T^3/45){\nu_h[1-\Theta(T-T_c)+\nu_q\Theta(T-T_c)}.
	\end{equation}
	\begin{equation}
		s^*(T)=s(T)/T^3,\ \ s^*_c=s_c/T^3_c,\ \ s_c=\frac{\pi^2T_c^3}{45}(\nu_h+\nu_q).
	\end{equation}
	
	The above simple bag model EOS can also be modified \cite{Kallmann, Sysoev} by making the bag "constant" $T$-dependent, $B(T)=AT$, to produce a metastable QGP state with negative temperature (see Fig.~\ref{Fig:6}c and Fig.~\ref{Fig:6}d):
	\begin{equation}
		p_q(T)=(\pi^2/90)\nu_qT^4-AT,\ \ \ p_h(T)=(\pi^2/90)\nu_hT^4.
	\end{equation}

	\begin{figure*}[ht]
		\centering
		\includegraphics[width=.94\textwidth]{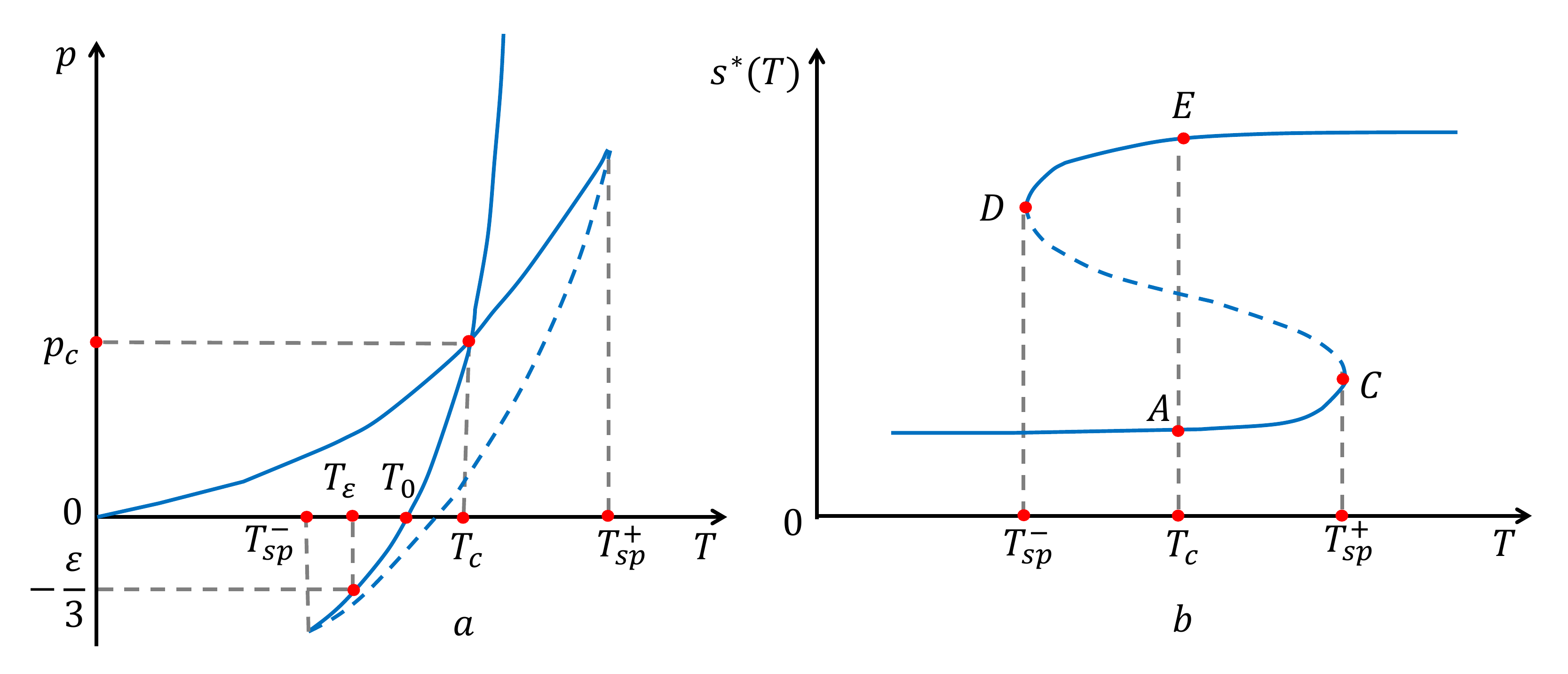}
		\caption{Modified bag EOS Eq. (\ref{EOS2}), including metastable states. Icon $a$ is the $p(T)$ dependence while icon $b$ is $s^*(T)=s(T)/T^3$.\label{Fig:Kall}}  
	\end{figure*}

	In Ref.~\cite{BO} EOS Eq.~(\ref{eq:EOS_ent1}) was modified by the replacement 
	\begin{equation}
		\Theta(T-T_c)\rightarrow \frac{1}{2}\left[1+\tanh \left(\frac{T-T_c}{\Delta T}\right)\right],
	\end{equation}
	where $\Delta T$ is a parameter related to the smoothness of a crossover transition. With the above substitution, the EOS can be rewritten as 
	\begin{equation}\label{Eq:EOS1} 
		\frac{T-T_c}{\Delta T}={\rm arctanh}(\Gamma\Delta s^*),
	\end{equation} 
	where $\Gamma=45/[\pi^2(\nu_q-\nu_h)],\ \ \Delta s^*=s^*-s^*_c$, behaving as
	shown by the dotted line in Fig.~\ref{Fig:6}b. This modification, however, smoothes down first order phase transitions in EOS, thus excluding possible metastable states. It may serve as a springboard for a further modification suggested in Ref. \cite{BJS}, namely: 
	\begin{equation} \label{EOS2} 
		(T-T_c)/\Delta T={\rm arctanh}(\Gamma\Delta s^*)-\gamma\Delta s^*,
	\end{equation}
	where $\gamma$ is the "metastability parameter". For $\Gamma-\gamma>0$, EOS Eq.~(\ref{EOS2}) is of the same type as Eq. (\ref{Eq:EOS1}), while for $\Gamma-\gamma<0$ a loop emerges in the entropy density (Fig.~\ref{Fig:Kall}b, similar to the loop in the density to pressure dependence, described by the Van der Waals equation or the EOS of a magnet in vicinity of the Curie point, replacing $\Delta s^*$ with the order parameter of the phase transition and substituting $(T-T_c)/\Delta T$ by the corresponding conjugate field. For $\Gamma-\gamma<0$ the curve $s^*(T)$ contains a non-physical region (interval $CD$) with $ds/dT<0$. The intervals ($AC$) and 
	($DE$) of $s^*(T)$ correspond to metastable states, $C$ and $D$ being the spinodal points where $ds/dT\sim c_s^{-2}$, with $c_s$ being the sound speed.

	Another important feature of the EOS Eq.~(\ref{EOS2}) is that for $\Gamma-\gamma=0$ it describes second order phase transitions, with singular behavior of the thermal capacity at $T=T_c.$ Really, in this case, we have $T-T_c\sim (\Delta s^*)^3$ near $T=T_c$.

	\section{Summary and conclusions}
	
	We related two different approaches to the critical point in the hadron spectra using statistical (Hagedorn) and dynamical (Regge) models. Our innovation is in the use of non-linear Regge trajectories, predicting a limited spectrum of hadron states, ignored in most of the papers on Regge-pole theory. Although our analyses is limited to the simple case of the $\rho$ meson spectrum, this technique and the results can (and will) be extended to other hadrons.

	In this work, we showed that the observed saturation of the number of hadrons as a function of mass can be easily explained in the Regge theory, by using non-linear trajectories and attributing the prefactor in Hagedorn's density to the slope of the corresponding Regge trajectory. Due to flattening of the latter, the slope vanishes, and the number of states flattens. In this way, we enter the most intriguing of the strong interaction, namely the expected transition of excited colorless hadrons into the hypothetical boiling soup of quarks and gluons.  
	
	Flattening of the exponential density of states and of the linear rise of Regge trajectories point to the same phenomenon, namely quark deconfinement (``melting" of hadrons). The two phenomena are correlated but they are not identical. Their combined study and further fits to the data may reduce the available freedom of the relevant parametrizations and tell us more about the unset of deconfinement. 
	
	The maximum mass of existing resonances depends on the parameters fitted to the observed resonances. 
	Resonances tend to disappear (fade) beyond some mass. Their non-observability may have two reasons: either they are not visible because of their large width (decay time) or they melt losing their individual features. These phenomena, especially the second, "boiling" phase can be studied also in the framework of thermodynamics, \textit{e.g.}, by studying the relevant equation of state (EoS). 
	

	\section*{Acknowledgements}
	
	We thank Wojciech Broniowski for useful correspondence on the subject of this paper.
	
	LJ was supported by the National Academy of Sciences of Ukraine's Grant N 1230/22-1
	"Fundamental properties of matter" and by the NKFIH Grant no. K133046; TB acknowledges support from the Hungarian National Office for Research, Development and Innovation, NKFIH under the project number K123815; ISz was supported the by the NKFIH Grant no. K133046 and by the \'UNKP-22-3 New National Excellence Program of the Ministry for Innovation and
	Technology from the source of the National Research, Development and Innovation Fund.

	
\end{document}